\documentclass[12pt]{article}
\usepackage[utf8]{inputenc}
\usepackage{amsmath, amssymb}
\usepackage{hyperref}
\usepackage{geometry}
\usepackage{parskip}
\usepackage{graphicx}
\usepackage{float}
\usepackage[style=apa, backend=biber]{biblatex} 
\title{Hierarchical Risk Parity for Portfolio Allocation in the Latin American NUAM Market}
\author{
    \makebox[.33\linewidth]{Gonzalo Ramírez-Carrillo} 
    \makebox[.33\linewidth]{David Ortiz-Mora} 
    \makebox[.33\linewidth]{Alex Aguilar-Larrotta} \\[1cm]
}

\date{}
\date{
    \small
    \begin{center}
        Faculty of Economics\\
        Rosario Investment Club\\
        Finance Research Group\\
        UNIVERSIDAD DEL ROSARIO\\
        Calle 12C No. 4-69\\
        Bogotá D.C., Colombia
    \end{center}
}
\begin{document}

\maketitle

\begin{abstract}
This study applies the Hierarchical Risk Parity (HRP) portfolio allocation methodology to the NUAM market, a regional holding that integrates the markets of Chile, Colombia, and Peru. As one of the first empirical analyses of HRP in this newly formed Latin American context, the paper addresses a gap in the literature on portfolio construction under cross-border, emerging market conditions. HRP leverages hierarchical clustering and recursive bisection to allocate risk in a manner that is both interpretable and robust—avoiding the need to invert the covariance matrix, a common limitation in traditional mean-variance optimization. Using daily data from 54 constituent stocks of the MSCI NUAM Index from 2019 to 2025, we compare the performance of HRP against two standard benchmarks: an equally weighted portfolio (1/N) and a maximum Sharpe ratio portfolio. Results show that while the Max Sharpe portfolio yields the highest returns, the HRP portfolio delivers a smoother risk-return profile, with lower drawdowns and tracking error. These findings highlight HRP’s potential as a practical and resilient asset allocation framework for investors operating in integrated, high-volatility markets like NUAM.
\end{abstract}

\section{Introduction}
The unified NUAM market, which integrates the stock exchanges of Chile, Colombia, and Peru, represents a fundamental step toward the integration of capital markets in Latin America \parencite{nuammercadounico}. By consolidating trading platforms and harmonizing regulatory frameworks, it aims to enhance cross-border liquidity, expand investor access, and stimulate capital flows among the three Andean economies. Due to its recent formation, this market remains underexplored in academic literature, particularly regarding portfolio optimization strategies tailored to its emerging and cross-border structure.

Modern Portfolio Theory (MPT), pioneered by Harry Markowitz \parencite{atmaca_portfolio_2022}, laid the groundwork for the mathematical construction of portfolios, focusing on the relationship between expected return and risk \parencite{markowitz_foundations_1991}. However, in practice, MPT has certain limitations, such as high sensitivity to errors in parameter estimation \parencite{sen_comparative_2022}, the need to invert the covariance matrix \parencite{salas-molina_empirical_2025}, and the concentration of asset weightings within the portfolio.

To address these challenges, \textcite{prado_building_2016} introduced the Hierarchical Risk Parity (HRP) methodology. HRP relies on the unsupervised machine learning technique known as hierarchical clustering (tree clustering) to construct diversified portfolios. This methodology improves out-of-sample performance and is a robust technique that avoids matrix inversion, a common issue in quadratic optimizers.

Several academic studies have demonstrated the versatility and robustness of the HRP methodology across various contexts. \textcite{sen_comparative_2022} analyzed 50 representative stocks from the Indian NIFTY 50 market using daily closing price data from 2016 to 2021. Their results indicate that HRP outperformed traditional approaches in terms of risk-adjusted returns, providing better diversification and lower portfolio volatility. Similarly, \textcite{salas-molina_empirical_2025} conducted an empirical evaluation of distance metrics in the HRP methodology using data from the Spanish IBEX 35, selecting three distinct periods of weekly returns: a bull market (2005–2008), a sideways trend (2014–2019), and a bear market (2008–2012). HRP maintained robustness across different market regimes, with notable improvements in downside risk control during bear markets. Likewise, \textcite{jothimani_risk_2019} investigated risk parity models, including HRP, in the Canadian market. Their study focused on a subset of 44 stocks from the TSX Composite Index (initially 60, filtered for complete data) covering 10 main sectors, using 10 years of data (January 2007 to December 2016). HRP achieved superior sectoral diversification and stable performance compared to equal-weighted and traditional risk parity strategies.

More recently, \textcite{millea_using_2023} applied the HRP methodology to the cryptocurrency market (117 assets), the stock market (46 assets), and the foreign exchange market (28 pairs). Their findings suggest that HRP, when combined with reinforcement learning, improved allocation efficiency and adapted dynamically to volatile and heterogeneous asset classes. Finally, \textcite{aragon_urrego_application_2023} examined a set of Latin American ADRs listed in the United States, comparing HRP’s performance against traditional asset allocation strategies. The results indicate that HRP provides greater diversification and improved risk control, particularly during periods of elevated market volatility.

Building on previous research, this study seeks to answer the following research question: Does a portfolio constructed using the Hierarchical Risk Parity (HRP) methodology offer greater diversification and more efficient risk management than traditional strategies such as equally weighted allocation (1/N) and maximum Sharpe ratio optimization, when applied to the assets that comprise the NUAM index?

In order to answer this question, the primary objective of the study is to apply the HRP methodology to the equity sample of the integrated NUAM market. The aim is to assess the ability of HRP to construct portfolios that are more resilient and superior in terms of diversification and efficiency. For this purpose, we use historical daily closing prices of 54 stocks from the MSCI NUAM Index, covering the period from July 2019 to June 2025. The results obtained from HRP are then compared against two benchmark strategies: an equally weighted portfolio (1/N) \parencite{pflug_1n_2012} and a maximum Sharpe ratio portfolio which is the combination of assets that produces the highest expected return per unit of risk taken \parencite{lo_statistics_2002}. Performance is evaluated through standard portfolio metrics such as annualized return, volatility, Sharpe ratio, Sortino ratio, Calmar ratio, maximum drawdown, and tracking error, providing a comprehensive risk–return assessment return \parencite{camacho_gestion_2023,bacon_practical_2008}.

The contribution of this research is threefold. First, it provides investors, portfolio managers, and financial institutions with a more robust alternative to traditional asset allocation frameworks, such as mean-variance optimization and the equally weighted portfolio, by empirically evaluating the Hierarchical Risk Parity (HRP) methodology. Second, it offers valuable quantitative evidence for regulators and market participants regarding relative performance and risk management in the context of the NUAM index. Finally, this study contributes to the academic literature by applying and analyzing HRP in an emerging Latin American market, which remains underexplored in the field of quantitative portfolio optimization.

The remainder of the paper is structured as follows: Section 2 describes the Hierarchical Risk Parity (HRP) methodology; Section 3 presents the dataset used in the study; Section 4 reports the empirical results; and Section 5 concludes with the main findings, limitations, and directions for future research.

\section{Methodology}

Modern Portfolio Theory (MPT), developed by Harry Markowitz in 1952, introduced a formal and quantitative framework for portfolio construction by modeling the trade-off between expected return and risk \parencite{markowitz_portfolio_1952}. In MPT, risk is represented by the variance (or standard deviation) of asset returns, while the expected return is calculated as a linear combination of individual asset returns weighted by their portfolio proportions \parencite{markowitz_foundations_1991}. The core insight of MPT is that portfolio-level risk can be reduced through diversification—combining assets that are not perfectly correlated—without necessarily sacrificing return.
The central objective in MPT is to solve the optimization problem:

\begin{equation}
\min_{w} \; w^\top \Sigma w \quad \text{subject to} \quad w^\top \mu = \mu_p,\; \sum_{i=1}^{n} w_i = 1,
\end{equation}

where \( w \) is the vector of asset weights, \( \Sigma \) is the covariance matrix of asset returns, and \( \mu \) is the vector of expected returns. This formulation seeks the minimum variance portfolio for a given level of expected return \( \mu_p \). The solution implies a closed-form expression:

\begin{equation}
w = \frac{\Sigma^{-1} \mu}{\mathbf{1}^\top \Sigma^{-1} \mu},
\end{equation}

This theory led to the formulation of the  efficient frontier, a set of portfolios that offer the highest expected return for a given level of risk. Portfolio selection is then reduced to solving a constrained quadratic optimization problem, relying on the expected return vector and the covariance matrix of asset returns \parencite{markowitz_portfolio_1952, marling_markowitz_nodate}. The mathematical elegance of the model, however, contrasts with the significant practical challenges it faces when implemented in real-world settings \parencite{kolm_60_2014}.

One of the most persistent limitations of MPT is its  sensitivity to estimation error \parencite{prado_building_2016}. In particular, errors in forecasting expected returns can lead to unstable and highly concentrated portfolios, a phenomenon widely observed across empirical studies and real-world applications \parencite{mena_valencia_modelo_2023}. Furthermore, the requirement to invert the covariance matrix can result in numerical instability, especially in high-dimensional problems where the number of assets exceeds or closely matches the length of historical data series, leading to portfolios that are overfitted and fragile out-of-sample \parencite{prado_building_2016}.

Moreover, MPT does not inherently control how risk is distributed across assets \parencite{palit_study_2024}. While it optimizes return per unit of total variance, it often produces portfolios where a few low-volatility assets dominate the allocation, undermining the notion of true risk diversification \parencite{ciciretti_network_2024}. This issue is particularly evident in emerging or integrated markets, where asset correlations and volatilities are more dynamic and less predictable.

To address these weaknesses, several extensions have been developed, including robust optimization, Bayesian methods, shrinkage estimators, and the Black–Litterman model \parencite{kaczmarek_building_2022}. These approaches aim to improve estimation stability or incorporate investor views, yet they still rely on the mean-variance framework and its structural assumptions \parencite{prado_building_2016}. Their effectiveness remains limited when market conditions are volatile, non-normal, or affected by structural breaks—as is often the case in frontier or emerging markets.

In recent years,  Hierarchical Risk Parity (HRP)  has emerged as a compelling alternative. Introduced by \textcite{prado_building_2016}, HRP is a non-parametric, data-driven method for portfolio allocation that avoids matrix inversion and does not require estimates of expected returns. This characteristic represents a significant advantage of the HRP method compared to traditional portfolio optimization approaches such as Markowitz’s Critical Line Algorithm (CLA) \parencite{salas-molina_empirical_2025}. HRP uses hierarchical clustering techniques to infer the structure of asset correlations and allocate risk recursively based on the tree structure. This results in more stable and interpretable portfolios, especially in scenarios with noisy data or high multicollinearity \parencite{daniel_aragon_urrego_paridad_2021}.

\subsection{HRP Algorithm Process}

The first step, tree clustering, is performed using an agglomerative hierarchical clustering algorithm \parencite{sen_comparative_2022}. This process involves grouping observations (stocks, in this study) with similar characteristics into clusters \parencite{diop_simrec_2025}, forming a hierarchical structure based on a selected distance metric. The procedure begins with each observation as an individual cluster and then iteratively merges the closest clusters until a single group encompassing all observations is formed \parencite{lu_unsupervised_2024}.

As an initial input for this step, a distance matrix is required \parencite{prado_advances_2018}. This matrix is built from the distances between each pair of observations. For this, an $(N \times N)$ correlation matrix of asset returns is used as a base, which is then transformed into distances as follows:

\[
d_{ij} = \sqrt{ \frac{1 - \rho_{ij}}{2} }
\]

Where \( d_{ij} \in [0;1] \) y \( \rho_{ij} \) is the correlation between assets $i$ and $j$.

The process of generating hierarchies requires a linkage method, such as single linkage, complete linkage, average linkage, or Ward’s method \parencite{yokokawa_validating_2022, habib_complex_2016}. To visually represent the resulting hierarchy from the clustering algorithm, a graph known as a dendrogram is used. This diagram displays the successive groupings, from individual elements to larger clusters \parencite{li_similarity-based_2024}.

Once the hierarchical structure has been established, the next step in the HRP method is the Quasi-diagonalization of the covariance matrix \parencite{prado_advances_2018}. This step involves reordering the rows and columns of the covariance matrix such that assets with higher covariances are placed closer to the main diagonal. Mathematically, this reordering is expressed as:

\[
C^* = P C P^T
\]

Where, $C^*$ is the quasi-diagonalized matrix, $C$ is the original covariance matrix $(N \times N)$ constructed from the historical returns of the assets under consideration, and $P$ is the permutation matrix $(N \times N)$ based on the hierarchical order resulting from the agglomerative clustering algorithm.

The final step of the HRP algorithm is Recursive Bisection \parencite{prado_advances_2018}, where allocations are assigned to each asset in the portfolio. This is a top-down process. Based on the order obtained during quasi-diagonalization, the algorithm recursively divides the ordered set of assets into two adjacent subsets. For each partition, the aggregated variance of both subsets is calculated, and a split factor $\alpha$ is determined, proportional to the inverse of their variances. The weights assigned to each subset are adjusted based on $\alpha$ and $1 - \alpha$, ensuring that the sum of all weights equals one:

\[
\alpha = \frac{\sigma^2_R}{\sigma^2_L + \sigma^2_R}, \quad w_L = \alpha \cdot w_s\ , \quad w_R = (1-\alpha) \cdot w_s
\]

where $\sigma^2_L$ and $\sigma^2_R$ are the variances of the left and right subsets, respectively, and $w_s$ is the total weight of the current subset before the division.

\subsection{Implementation and Specific Parameters of the Study}

The computational implementation of the Hierarchical Risk Parity (HRP) algorithm was conducted in Python, following the methodology proposed by \textcite{prado_advances_2018}. The algorithm is structured in three key stages: (1) Tree Clustering, using a distance metric derived from asset correlations; (2) Quasi-diagonalization of the covariance matrix based on the hierarchical structure; and (3) Recursive Bisection for weight allocation, grounded on the variance of subclusters.

As part of the Tree Clustering process, the dissimilarity between assets was measured using the transformed correlation metric $\sqrt{0.5(1 - \rho)}$. Asset returns were computed as continuously compounded log returns using daily closing prices.

Additionally, a comparative evaluation was conducted against two traditional portfolio strategies: an equally weighted portfolio, defined as $w = \frac{1}{N}$, and a maximum Sharpe ratio portfolio. This comparison was based on widely accepted performance metrics, including annualized return, annualized volatility, Sharpe Ratio, Sortino Ratio, Calmar Ratio, maximum drawdown, and tracking error \parencite{kaczmarek_building_2022}.

\section{Data}
As part of the integration of the Chilean, Colombian, and Peruvian stock exchanges under the NUAM market, the MSCI NUAM Index was formed to represent the most liquid and representative equities across these markets. As of april 2025, the Index includes 54 stocks: 33 from Chile, 13 from Colombia, and 8 from Peru. This composition reflects capitalization and liquidity criteria applied quarterly by MSCI to ensure that only the most actively traded and financially relevant companies are included. 

To investigate the application of the Hierarchical Risk Parity (HRP) methodology to the integrated NUAM market, we retrieved historical daily closing prices for the 54 constituent stocks of the MSCI NUAM Index, the data was sourced from the Bloomberg Terminal, a trusted platform for financial market data, ensuring accuracy and reliability. The dataset spans from July 2019, to June 2025, encompassing pre-pandemic market conditions (2019–2020), the volatile period of the COVID-19 pandemic (2020–2021), and the subsequent recovery and operational phase of the NUAM market. This six-year period was selected to capture diverse market dynamics, enabling a robust evaluation of HRP’s performance across stable, turbulent, and transitional economic environments. The index includes companies from sectors such as Financials, Materials, Consumer Staples, and Utilities, ensuring comprehensive market representation as detailed in Table 1.

\begin{table}[H]
\centering
\caption{Number of Stocks per Sector in the Nuam MSCI Index}
\vspace{0.5em}
\footnotesize
\begin{tabular}{lc}
\hline
\textbf{Sector} & \textbf{Number of Stocks on the Index} \\
Communication Services      & 1  \\
Consumer Discretionary      & 4  \\
Consumer Staples            & 6  \\
Energy                      & 1  \\
Financials                  & 15 \\
Industrials                 & 6  \\
Information Technology      & 1  \\
Materials                   & 8  \\
Real Estate                 & 4  \\
Utilities                   & 8  \\
\hline
\end{tabular}

\end{table}

This comprehensive dataset provides a solid foundation for analyzing the effectiveness of HRP in building diversified and resilient portfolios within the emerging, cross-border NUAM market context. 
\section{Results}
This section presents the empirical findings derived from the application of the HRP methodology to the selected stock sample. The analysis is structured to evaluate HRP's ability to generate more diversified portfolios compared to an equally weighted portfolio.

Figure 1 displays the dendrogram built from the dissimilarity matrix between assets. This dendrogram graphically represents the hierarchical clustering structure of the investment universe assets and serves as the foundation for the Hierarchical Risk Parity (HRP) algorithm.

The horizontal axis represents individual assets, while the vertical axis shows the dissimilarity distance at which they are grouped. Each branch represents a cluster merger: the higher the junction, the more distinct the merged groups were. By cutting the dendrogram at a height of 1, three well-defined main clusters can be identified, each colored in a different shade. These groups reflect subsets of assets with high internal correlation and low inter-cluster correlation, which is key for achieving better portfolio diversification.

This hierarchical pattern is exploited by HRP to assign weights efficiently: more similar assets receive differentiated weights within each cluster, and more dissimilar clusters are assigned weights inversely proportional to the group’s aggregate risk, following the logic of recursive bisection.

\begin{figure}[H]
    \centering
    \includegraphics[width=0.87\linewidth]{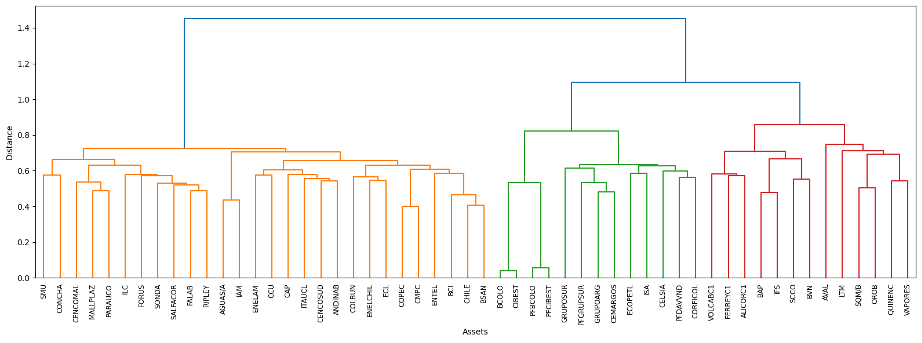}
    \caption{Hierarchical Clustering Dendrogram of Assets}
    \label{fig:enter-label}
\end{figure}

Figures 2 and 3 present two heatmaps that illustrate the correlation structure between assets before and after applying hierarchical reordering through the HRP algorithm. Figure 2 corresponds to the dissimilarity matrix based on the original correlations. The intensity of the red color suggests a dense network of correlations among assets, without an apparent structure.

\begin{figure}[H]
    \centering
    \includegraphics[width=0.6\linewidth]{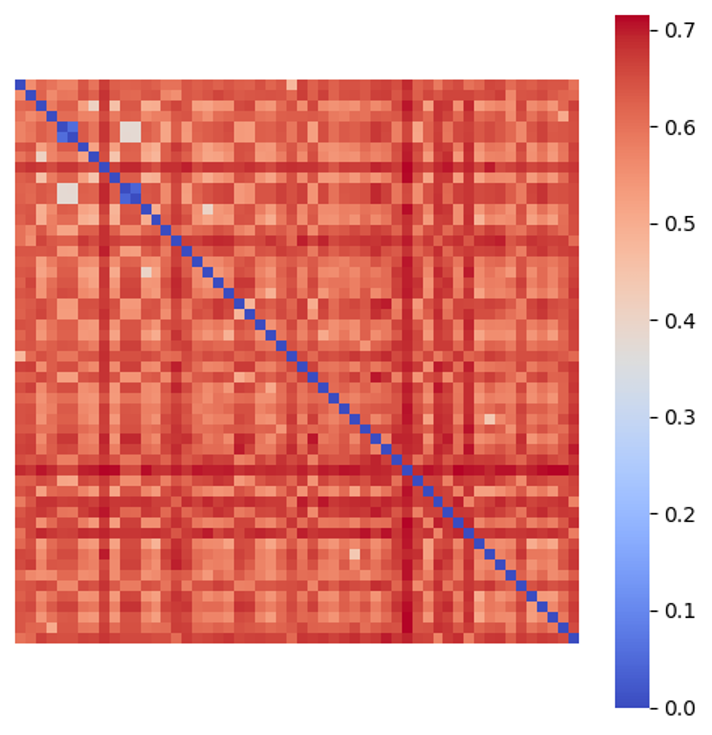}
    \caption{Correlation-Based Distance Matrix}
    \label{fig:enter-label}
\end{figure}

In contrast, Figure 3 shows the quasi-diagonalized matrix. A clearer organization is evident: the blue blocks concentrated near the main diagonal indicate groups of assets with high internal correlation and low correlation across groups. This reordering not only enhances visualization but also allows the HRP algorithm to exploit the hierarchical structure of asset relationships to assign weights more efficiently and robustly.

\begin{figure}[H]
    \centering
    \includegraphics[width=0.61\linewidth]{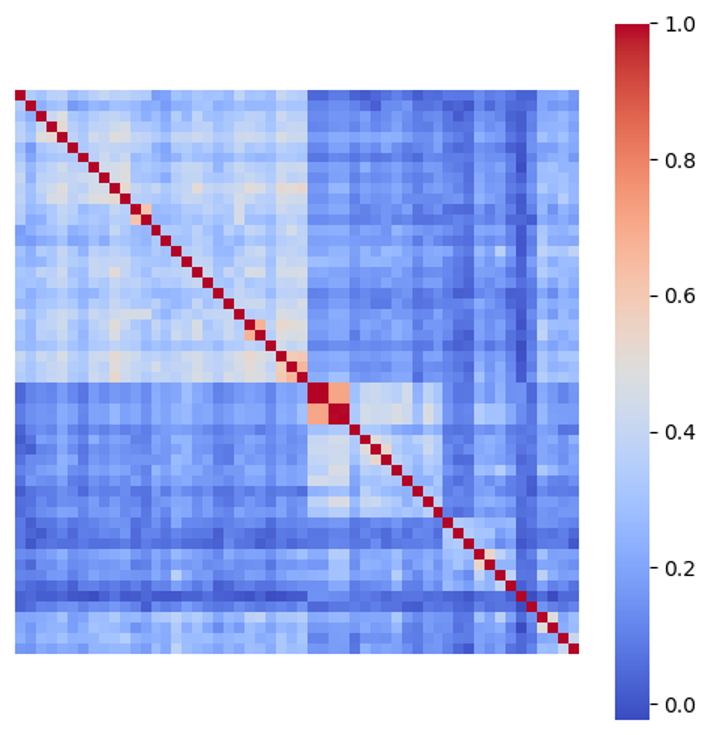}
    \caption{Quasi-Diagonalized Covariance Matrix}
    \label{fig:enter-label}
\end{figure}

Figure 4 shows the weight allocation resulting from the HRP algorithm, sorted from highest to lowest. This graphical representation highlights the non-uniform weight distribution, with a clear preference for assets belonging to less correlated clusters. Among the assets with the highest weights are AVAL (5.83\%), CORFICOL (4.89\%), FERREYC1 (3.71\%), and ALICORC1 (3.36\%). These assets likely belong to less risky or well-diversified clusters and therefore receive higher allocations. Conversely, assets such as LTM, GRUPOSUR, and CIBEST have the lowest weights, possibly due to higher risk concentration or strong correlation with other assets in the same group.

\begin{figure}[H]
    \centering
    \includegraphics[width=0.98\linewidth]{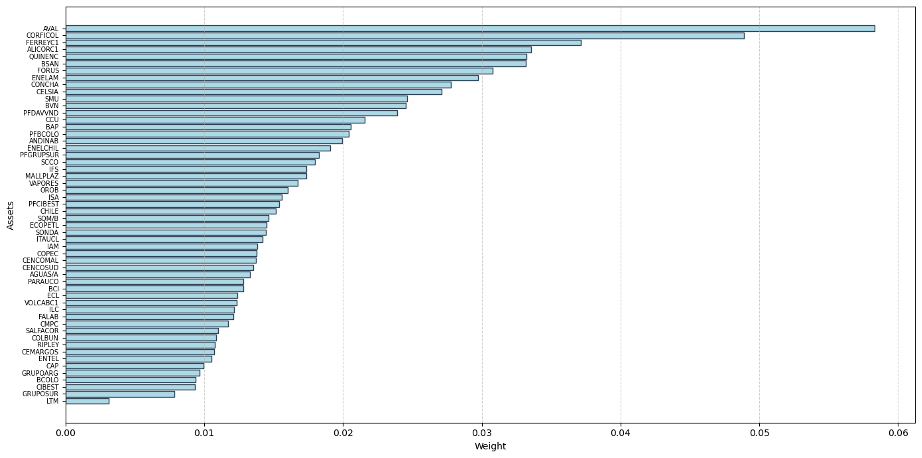}
    \caption{Asset Weights According to the HRP Method}
    \label{fig:enter-label}
\end{figure}

Figure 5 presents the cumulative returns of the three portfolios analyzed: the Hierarchical Risk Parity (HRP) portfolio, the Maximum Sharpe portfolio, and the equally weighted benchmark portfolio (1/N). As shown, the Maximum Sharpe portfolio achieves higher cumulative returns but also exhibits greater volatility. On the other hand, the HRP portfolio offers a more stable performance, with a smoother wealth accumulation. These patterns reinforce the idea that HRP is particularly well-suited for investors seeking a diversification framework focused on risk control, while the Maximum Sharpe approach may be more attractive to return-maximizing profiles with higher risk tolerance.
\begin{figure}[H]
    \centering
    \includegraphics[width=0.98\linewidth]{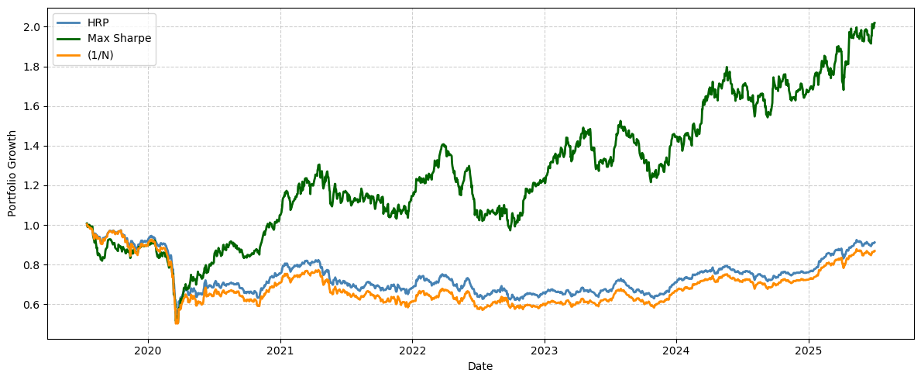}
    \caption{Cumulative Returns of the Portfolios}
    \label{fig:enter-label}
\end{figure}

Table 2 presents the performance and risk metrics for the three evaluated portfolios: HRP, Maximum Sharpe, and the equally weighted (1/N) portfolio. The results reveal a significant contrast between the strategies.

The Max Sharpe portfolio exhibits the best overall performance, with an annualized return of 14.2\%, a Sharpe ratio of 0.426, and the highest Sortino ratio (0.529), suggesting favorable risk-adjusted returns, even when considering only downside volatility. This portfolio also presents the highest Calmar ratio (0.205), indicating an acceptable balance between return and extreme drawdown.

In contrast, both the HRP and the equally weighted portfolios show negative returns (--0.3\% and --0.7\%, respectively) and similar Sharpe ratios (--0.278 and --0.274), indicating that neither managed to overcome the volatility associated with the returns during the analyzed period. However, the HRP portfolio displayed a slightly lower maximum drawdown and a reduced tracking error (0.037), reflecting better relative management of systematic risk.

\begin{table}[H]
\centering
\caption{Comparative Performance Metrics of Portfolios}
\vspace{0.5em}
\resizebox{\textwidth}{!}{%
\begin{tabular}{lccccccc}
\hline
\textbf{Portfolio} & \textbf{Annual Return} & \textbf{Volatility} & \textbf{Sharpe} & \textbf{Sortino} & \textbf{Calmar} & \textbf{Max Drawdown} & \textbf{Tracking Error} \\
HRP         & -0.003 & 0.155 & -0.278 & -0.299 & -0.096 & -0.452 & 0.037 \\
Max Sharpe  &  0.142 & 0.239 &  0.426 &  0.529 &  0.205 & -0.496 & 0.164 \\
1/N         & -0.007 & 0.175 & -0.274 & -0.293 & -0.096 & -0.500 & --    \\
\hline
\end{tabular}
}
\end{table}

\section{Discussion \&\ Conclusions}
Portfolio Optimization is an important topic in finance worldwide. As new markets emerge, so do opportunities along investors looking for innovative forms of diversifying asset allocation. The NUAM market, unifying the stock exchanges of Colombia, Chile, and Peru, represents a transformative step toward an integrated Latin American capital markets. This study is among the first to explore Hierarchical Risk Parity (HRP) portfolio optimization in this emerging financial ecosystem, highlighting its potential as a robust strategy for investors. By employing hierarchical clustering and recursive risk allocation, HRP constructs portfolios that outperforms the equally weighted benchmark in managing systematic risk, offering a smoother wealth accumulation path. However, compared to a maximum return-focused strategy, HRP prioritizes risk control over aggressive gains, appealing to investors seeking stability in a volatile market, a common characteristic in emerging markets.

 These findings offer practical insights for investors aiming to diversify portfolios but also for regulators evaluating the efficiency, stability and impact of NUAM’s integrated framework. 
As the Andean market evolves, further research into its unique economic drivers and extended post-integration data will enhance HRP’s potential, driving innovative approaches to portfolio management in emerging markets.

Future research could incorporate dynamic correlation measures to better capture the evolving relationships among assets in volatile environments. It would also be relevant to explore variants of the HRP algorithm that integrate macroeconomic, sectoral, or other relevant dimensions into asset selection. Additionally, traditional distance matrices could be combined with alternative representations that capture additional asset characteristics, potentially enhancing the clustering process and improving portfolio diversification.

\printbibliography

\end{document}